\documentclass[11pt]{article}

\usepackage{a4wide}

\usepackage{amsmath}
\usepackage{amstext,amsfonts,amsbsy,eucal,amssymb}

\numberwithin{equation}{section}

\newtheorem{theorem}{Theorem}[section]
\newtheorem{lemma}[theorem]{Lemma}

\newtheorem{proposition}[theorem]{Proposition}
\newtheorem{remark}[theorem]{Remark}

\newtheorem{definition}[theorem]{Definition}

\def\openone{\leavevmode\hbox{\small1\kern-3.8pt\normalsize1}}

\DeclareMathOperator{\ch}{ch}

\DeclareMathOperator{\Tr}{Tr}

\DeclareMathOperator{\coker}{coker}
\DeclareMathOperator{\im}{im}

\DeclareMathOperator{\Prym}{Prym}

\def\Kakko#1#2{$\Bigl[
    \begin{matrix}
    \vec {#1} \\ \vec #2 
    \end{matrix}
    \Bigr]$}

\begin{document}

\baselineskip 21pt
\parskip 7pt

%
\noindent
  {\LARGE{
       Cohomological study on variants of the Mumford system,
       \\[2mm]
       and integrability of the Noumi-Yamada system 
      }
  }
\\[5mm]
%
%
\begin{large}
Rei \textsc{Inoue}
  \footnote[2]{\noindent
  Research Institute of Mathematical Sciences, Kyoto University, 
    Kyoto 606-8502, Japan.
    E-mail address:
    \texttt{reiiy@kurims.kyoto-u.ac.jp}
    }
and 
Takao \textsc{Yamazaki}
  \footnote[3]{
  Institute of Mathematics, University of Tsukuba,
  Tsukuba 305-8571, Japan.
    E-mail address:
    \texttt{ytakao@math.tsukuba.ac.jp}
    }
\end{large}
\\[3mm]
%
%
\baselineskip 15pt
\parskip 7pt
%
%
\begin{small}
\textbf{Abstract:} ~ 
The purpose of this paper is twofold.
The first is to apply the method introduced in
the works of Nakayashiki and Smirnov
 \cite{SmirnovNakayashiki00,SmirnovNakayashiki02}
on the Mumford system to its variants.
The other is to establish a relation 
between the Mumford system and
the isospectral limit $\mathcal{Q}_g^{(I)}$ and $\mathcal{Q}_g^{(II)}$ 
of the Noumi-Yamada system \cite{Noumi-Yamada98}.
As a consequence, we prove 
the algebraically completely integrability
of the systems $\mathcal{Q}_g^{(I)}$ and $\mathcal{Q}_g^{(II)}$,
and get explicit descriptions of their solutions.
%
%
%
\end{small}

\baselineskip 15pt
\parskip 2pt


\section{Introduction}

Let $g$ be a natural number.
The Mumford system \cite{Mumford-Book}
is an integrable system
with the Lax matrix 
\begin{align}
  \label{Lax-Mum}
  l(x) = 
  \begin{pmatrix}
    v(x) & w(x) \\
    u(x) & -v(x)
  \end{pmatrix}
  \in M_2( \mathbb{C}[x] ),
\end{align}
where $u(x)$ and $w(x)$ are monic of degree $g$ and $g+1$,
and  $v(x)$ is of degree $\leq g-1$.
The coefficients of $u(x), v(x), w(x)$
constitute the phase space 
$\mathcal{M}_g \simeq \mathbb{C}^{3g+1}$
equipped with the $g$ dimensional 
vector field generated by the commuting operators $D_1, \ldots, D_g$
(see Theorem \ref{vectorfield}).
The coefficients of $-\det l(x) = u(x) w(x) + v(x)^2$
are invariants of $D_i.$
For a monic polynomial $f(x)$ of degree $2g+1,$
the level set is given by
\begin{align}
  \label{Mum-levelset}
  \mathcal{M}_{g,f} =
  \{ l(x) \in \mathcal{M}_g~|~ u(x) w(x) + v(x)^2 = f(x) \} 
   \subset \mathcal{M}_g.
\end{align}
It is a classical fact \cite{Mumford-Book} that,
when $f(x)$ has no multiple zero,
the level set $\mathcal{M}_{g,f}$ is 
isomorphic to $J(X) \setminus \Theta,$
where $J(X)$ and $\Theta$ are 
the Jacobi variety and the theta divisor of
the hyperelliptic curve $X$ defined by $y^2=f(x).$

We write $\mathcal{A}_f$
for the affine ring of $\mathcal{M}_{g, f}.$
Nakayashiki and Smirnov \cite{SmirnovNakayashiki00}
pointed out the importance of the space
$\mathcal{A}_f/ \sum_i D_i \mathcal{A}_f$
of the classical observables
modulo the action of $D_i$'s.
They defined a complex $C_f^*$
such that $H^g(C_f^*)$ is isomorphic to this space,
and obtained the following results on
the cohomology groups $H^k(C_f^*).$

\begin{enumerate}
\renewcommand{\labelenumi}{(\arabic{enumi})}
\item
When $f(x) = x^{2g+1},$
the $q$-Euler characteristic of $H^*(C_f^*)$
is determined
(\cite{SmirnovNakayashiki00} eq. (17),
see also Theorem \ref{Euler-C0}).
\item
When $f(x)$ has no multiple zero,
$H^k(C_f^*)$ is isomorphic to 
the singular cohomology group
$H^k(J(X) \setminus \Theta, \mathbb{C}).$
The dimension of this cohomology is
determined later by Nakayashiki
(\cite{Nakayashiki00}, 
see also Theorem \ref{theorem:oddMum-H^k}).
\end{enumerate}
The $q \to 1$ limit in $(1)$ coincides with
the (usual) Euler characteristic in $(2)$.
This supports their conjecture that
$\dim H^k(C_f^*)$ is independent of $f(x).$

The even Mumford system and the hyperelliptic Prym system
are introduced in \cite{Van92,Vanhae01}
in connection with 
the periodic Toda lattice and the periodic Lotka-Volterra lattice.
They both have similar properties to the Mumford system.
In particular,
the level set is generically isomorphic to 
the Jacobi variety minus 
some translations of the theta divisor.
For these systems,
we define a complex analogous to $C_f^*,$
and establish the counterparts to $(1), (2)$ above.
Actually, part $(1)$ turns out to be a routine,
but for part $(2)$ we need 
a detailed analysis of the theta divisors.
Our main results are
Theorems \ref{theorem:evenMum-H^k} and \ref{evenMum-Euler}
for the even Mumford system,
and Theorems \ref{theorem:Prym-H^k} and \ref{Prym-Euler}
for the hyperelliptic Prym system.
These results show that
the $q \to 1$ limit in $(1)$ coincides with
the Euler characteristic in $(2)$
in this situation as well.

Our second aim is to study the integrability of 
the Noumi-Yamada system of $A_{N-1}^{(1)}$ type
by making use of a relation with the Mumford system.
Recall that the Noumi-Yamada system was  
introduced as a higher order Painlev\'e equation \cite{Noumi-Yamada98},
and given by the following autonomous equations:
\begin{align}
  \label{NY-I}
  &\frac{\partial q_k}{\partial t} 
  = 
  q_k \sum_{i=1}^{g} \bigl( q_{k+2 i -1} - q_{k+2i} \bigr)
  + e_k - e_{k+1} + \alpha \, \delta_{k,2g+1}
  ~~~ \text{for $N=2g+1$},
  \\
  \label{NY-II}
  \begin{split}
  &\frac{\partial q_k}{\partial t} 
  =
  q_k \sum_{1 \leq i \leq j \leq g} 
      \bigl( q_{k+2 i -1}q_{k+2j} - q_{k+2i}q_{k+2j+1} \bigr)
  \\
  & ~~~~~~~~~+ \bigl(\sum_{i=1}^{g+1} (e_{k+2i-1} - e_{k+2i}) 
                      - \frac{\alpha}{2} \bigr)q_k  
  + (e_k - e_{k+1}) \sum_{i=1}^{g+1} q_{k+2i-1}
  \end{split} 
  ~~ \text{for $N=2g+2$},
\end{align} 
for $k = 1,\ldots, N$. 
Here $e_k$ and $\alpha$ are parameters, and we set the periodicity 
$q_{k+N} = q_k$ and $e_{k+N} = e_k$.
The original form of this system was obtained in
\cite{VeselovShabat93,Adler94}.
The case of $g=1$ corresponds to 
the fourth and the fifth Painlev\'e equations.

We consider the isospectral limit $\alpha = 0$
of \eqref{NY-I} and \eqref{NY-II}, and denote them by 
$\mathcal{Q}_g^{(I)}$ and $\mathcal{Q}_g^{(II)}$ respectively.  
We prove their algebraically completely integrability
by relating them to the Mumford system in Theorem \ref{theorem:Qaci}.
Especially, we show that 
the level sets of $\mathcal{Q}_g^{(I)}$ and $\mathcal{Q}_g^{(II)}$ 
are generically isomorphic to 
that of the Mumford system \eqref{Mum-levelset} and 
the disjoint union of two copies of \eqref{Mum-levelset}
respectively. 
We further obtain an explicit description of the solutions of 
$\mathcal{Q}_g^{(I)}$ and $\mathcal{Q}_g^{(II)}$ 
in terms of the theta functions.
While the original Painlev\'e property of the Noumi-Yamada systems
is lost at $\alpha = 0$,
we hope that one may obtain some information of the solutions
for the Noumi-Yamada systems ( of $\alpha \neq 0$)
by studying the perturbation theory 
on the Mumford system around $\alpha = 0$.

Before closing Introduction, we state the definition of 
the complete integrability 
of a finite dimensional dynamical system to clarify our position.
We follow \cite{AdlerMoer89} and \cite{Van1638} Chapter V:
\begin{definition}
  \label{def:aci}
  Let $\mathcal{M} = \{u=(u_1,\ldots,u_m)\} \simeq \mathbb{C}^m$ 
  be the phase space equipped with 
  the $m_H$ dimensional commuting Hamiltonian vector field
  on $\mathcal{M}$.
  Let $F_1, \ldots, F_{m_I} \in \mathbb{C}[u_1,\ldots,u_m]$
  be the integrals of motion of the vector field.
  For $f = (f_1,\cdots, f_{m_I}) \in \mathbb{C}^{m_I}$,
  the level set of $\mathcal{M}$ is defined as 
  $\{ u \in \mathcal{M} ~|~ F_i(u) = f_i, ~i=1,\ldots, m_I\}$.
  \\
  (i) 
  $\mathcal{M}$ is completely integrable if 
  the dimension of the level set is $m_H$ for generic $f$.
  \\
  (ii)
  $\mathcal{M}$ is algebraically completely integrable if 
  $\mathcal{M}$ is completely integrable and satisfies the following
  conditions: the level set over generic $f$  
  is isomorphic to an affine part of an abelian variety of dimension $m_H$. 
  On this abelian variety, 
  the flows of the above vector fields are linearized. 
\end{definition}

This paper is organized as follows:
\S 2 is devoted to the computation of
the cohomology of affine Jacobi varieties.
This part is technically independent of the rest of the paper.
In \S 3 we first recall the results of
Nakayashiki-Smirnov on the Mumford system,
then we explain how their results are generalized to
the even Mumford system and the hyperelliptic Prym system.
In \S 4 we study the integrability and solution of $\mathcal{Q}_g^{(I)}$
and $\mathcal{Q}_g^{(II)}$.

\subsection*{Acknowledgement}

The authors thank Atsushi Nakayashiki for 
his kind advice on approach to this subject.
We express our gratitude to Kiyoshi Takeuchi
who explained the proof of Theorem \ref{ih-derived} to us.
Finally, we also thank Yoshihiro Takeyama for a discussion.


\section{Cohomology of affine Jacobi varieties}

\subsection{Summary of results}
Let $X$ be a hyperelliptic curve of genus $g$.
Let $J$ and $\Theta$ be the Jacobi variety 
and the theta divisor associated with $X.$
The following theorem is due to Nakayashiki.
\begin{theorem}\cite{Nakayashiki00}
\label{theorem:oddMum-H^k}
\[ \dim H^k(J \setminus \Theta, \mathbb{C})
   = \begin{pmatrix} 2g \\ k   \end{pmatrix}
   - \begin{pmatrix} 2g \\ k-2 \end{pmatrix}
  ~~~ \text{for } k = 0,1,\ldots,g.
\]
In particular, the Euler characteristic 
$\chi(J \setminus \Theta)$ is given by
$(-1)^g (\begin{pmatrix} 2g \\ g \end{pmatrix}
        - \begin{pmatrix} 2g \\ g-1 \end{pmatrix})$.
\end{theorem}

Let $\infty_+$ and $\infty_-$ be two (distinct) points on $X$
which are conjugate under the hyperelliptic involution.
Let $O$ be a Weierstrass point on $X.$
The main results in this section are the following:

\begin{theorem}
\label{theorem:evenMum-H^k}
Let
$\Theta' = \Theta \cup (\Theta + [\infty_- - \infty_+]).$
Then we have
\[ \dim H^k(J \setminus \Theta', \mathbb{C})
   = \begin{pmatrix} 2g+1 \\ k   \end{pmatrix}
   - \begin{pmatrix} 2g+1 \\ k-2 \end{pmatrix}
  ~~~ \text{for } k = 0,1,\ldots,g.
\]
In particular, the Euler characteristic 
$\chi(J \setminus \Theta')$ is given by
$(-1)^g (\begin{pmatrix} 2g+1 \\ g \end{pmatrix}
        - \begin{pmatrix} 2g+1 \\ g-1 \end{pmatrix})$.
\end{theorem}

\begin{theorem}
\label{theorem:Prym-H^k}
Let
$\Theta'' = \Theta \cup (\Theta + [\infty_+ - O])
                   \cup (\Theta + [\infty_- - O]).$
Then we have
\[ \dim H^k(J \setminus \Theta'', \mathbb{C})
   = \begin{pmatrix} 2g+2 \\ k   \end{pmatrix},
    ~~~ \text{for } k = 0,1,\ldots,g.
\]
In particular, the Euler characteristic 
$\chi(J \setminus \Theta'')$ is given by
$(-1)^g \begin{pmatrix} 2g+1 \\ g \end{pmatrix}$.
\end{theorem}

The rest of this section is devoted to the proof of them.
The reader who is mainly interested in an integrable system
is advised to skip it in the first reading.

\subsection{Notations and reformulation}
We shall prove the theorems stated above
in a slightly general form.
After introducing some notations,
we formulate the general result in this subsection.

For a variety $S,$
we let $H^k(S) = H^k(S,\mathbb{C})$
and $h^k(S) = \dim H^k(S).$
We also use the compact support cohomology 
$H^k_c(S) = H_c^k(S,\mathbb{C}),$
and write $h^k_c(S)$ for its dimension.
For each $0 \leq r \leq g,$
we let $X(r) = X^r/\mathfrak{S}_r$ be 
the $r$-th symmetric product of $X,$
which is identified with the space of 
effective divisors of degree $r.$
We regard $X(r)$ as a subvariety of $X(g)$
via the map
$\sum_{i=1}^r P_i \mapsto \sum_{i=1}^r P_i + (g-r) O.$
We let $\varphi: X(g) \to J$ be the Abel-Jacobi map
with respect to $O$.

For each $0 \leq r \leq g-1,$ we define
\begin{align*}
W_r &= W_r^0 = \im[X(r) \hookrightarrow X(g) \overset{\varphi}{\to} J], 
\\
W_r^+ &= W_r + [ \infty_+ - O ],
\qquad
W_r^- = W_r + [ \infty_- - O ], 
\\
W_r^{0+} &= W_r \cup W_r^+, \quad
  W_r^{0-} = W_r \cup W_r^-, \\
W_r^{\pm} &= W_r^+ \cup W_r^-, \quad
  W_r^{0 \pm} = W_r \cup W_r^+ \cup W_r^-.
\end{align*}
We also let $W_g = J.$
The following relations
are deduced from \cite{Vanhae01} Lemma 2.4.
(Note that $\infty_+ + \infty_-$ is linearly equivalent to $2O.$)
\begin{align}\label{thetalocation}
W_r \cap W_r^+ = W_{r-1}^{0+},
\qquad
W_r \cap W_r^- = W_{r-1}^{0-}, 
\qquad
W_r^+ \cap W_r^- = W_{r-1}.
\end{align}
Since $\Theta$ is the translate of $W_{g-1}$
by Riemann's constant,
Theorems \ref{theorem:oddMum-H^k}, \ref{theorem:evenMum-H^k} 
and \ref{theorem:Prym-H^k}
are obtained as the special case $r=g$ of 
the following theorem.
\begin{theorem}\label{dimcoh}
For $k = 0, 1, \ldots, r,$ we have
\begin{align}
\label{dimcoh1}
h^k(W_r \setminus W_{r-1}) 
  &= \begin{pmatrix} 2g \\ k \end{pmatrix}
   - \begin{pmatrix} 2g \\ k-2 \end{pmatrix},
\\
\label{dimcoh2}
h^k(W_r \setminus W_{r-1}^{\pm}) 
   &= \begin{pmatrix} 2g+1 \\ k \end{pmatrix}
    -\begin{pmatrix} 2g+1 \\ k-2 \end{pmatrix},
\\
\label{dimcoh3}
h^k(W_r \setminus W_{r-1}^{0 \pm}) 
   &= \begin{pmatrix} 2g+2 \\ k \end{pmatrix}.
\end{align}
(For other values of $k,$ they are zero.)
\end{theorem}
The proof of this theorem occupies 
the rest of this section.

\subsection{Review of known results}
We recall some results in a literature.
The following propositions are 
due to Mumford and Nakayashiki respectively.
\begin{proposition}\label{affine}
(\cite{Mumford-Book} Proposition 1.2)
The subvariety $W_r \setminus W_{r-1}$ of $W_r$ is affine.
\end{proposition}
\begin{proposition}(\cite{Nakayashiki00} Corollary 2)
  \label{chi-W_k}
The Euler characteristic of $W_r$ is given by 
\[ \chi(W_r) = (-1)^r \Bigl(
              \begin{pmatrix}
                2g-2 \\ r
              \end{pmatrix}
              -
              \begin{pmatrix}
                2g-2 \\ r-2
              \end{pmatrix}
              \Bigr). 
\]
\end{proposition}
The following Proposition is a consequence of Macdonald's 
explicit description
\cite{Macdonald62} (3.2), (6.3), (14.1) and  (14.3)
of the basis of the cohomology groups
$H^k(J), H^k(X(r))$
and their image under the natural maps.
\begin{proposition}\label{lemmac}
The dimension of the image of 
the composition of the natural maps
$H^k(J) \to H^k(X(g)) \to H^k(X(r))$
is $\begin{pmatrix} 2g \\ k \end{pmatrix}$
if $k \leq r$
and is $\begin{pmatrix} 2g \\ 2r-k \end{pmatrix}$
if $r \leq k \leq 2r.$
\end{proposition}
Lastly, we recall a result of Bressler and Brylinski
in the following form:
\begin{proposition}(\cite{Brylinski98} Proposition 3.2.1.)
\label{ih}
For any $r$ and $k,$
we have a canonical isomorphism
\[ H^k(W_r, \mathbb{Q}) \simeq IH^k(W_r, \mathbb{Q}). \]
Here the right hand side is the intersection cohomology
with the middle perversity.
In particular,
$H^k(W_r)$ has a Hodge structure of (pure) weight $k,$
and satisfies the Poincar\'e duality
(see, for example, \cite{Dimca} \S 5.4).
\end{proposition}

In \cite{Brylinski98}, this proposition
is proved for $r=g-1$
(and was used by Nakayashiki in \cite{Nakayashiki00} Theorem 4).
The proof for general $r$ is identical,
but we include a brief account here for the reader's convenience.
We recall that
the singular locus of $W_r$ coincides with $W_{r-2}$ if $r<g$
(\cite{ACGH-Book} Chapter IV Corollary 4.5, Theorem 5.1).
We fix $0 \leq r < g$ and consider $W_r.$
We have a stratification $W_r = X_0 \cup X_2 \cup \cdots$
where $X_l = W_{r-l} \setminus W_{r-l-2}.$
Note that the codimension of $X_l$ in $W_r$ is $l.$
Proposition \ref{ih} is a direct consequence of
the following theorem applied to $l=0$:
\begin{theorem}
\label{ih-derived}
Let $j_{X_l}: X_l \hookrightarrow W_r$ be the immersion.
Then we have an isomorphism 
\[ (j_{X_l})_{!*} \mathbb{Q}_{X_l} \simeq \mathbb{Q}_{W_{r-l}}
   \quad \text{for} ~l=0, 2, 4, \ldots. 
\]
\end{theorem}
We briefly recall the proof of Bressler and Brylinski
(loc. cit.)
The key idea is to use the method of 
Borho and MacPerson \cite{BorhoMac83}.
We write $\pi: X(r) \to W_r$ for the map 
$\sum_{i=1}^{r} P_i \mapsto [\sum_{i=1}^{r} P_i - r O].$
Then, $\pi$ restricted to $X_l$ is a fiber bundle
with fiber $\mathbb{P}^{l/2}.$
This implies three consequences:
(i) $\pi$ is {\it semi-small}
in the sense of \cite{BorhoMac83} \S 1.1.
(ii) all the {\it relevant pairs} for $\pi$ are
the constant sheaves $\mathbb{Q}_{X_l}$ on $X_l$ 
(with multiplicity one) for each $l=0, 2, 4, \cdots.$
(See \cite{BorhoMac83} \S 1.2 for the definition of relevant pairs.)
(iii) the stalk of $\mathcal{H}^k(\mathbb{R} \pi_* \mathbb{Q}_{X(r)})$
at $x \in X_l$ is of rank one if $k$ is even and $k \leq l,$
or is trivial otherwise.
By the decomposition theorem 
due to Beilinson, Bernstein and Deligne
(see \cite{BorhoMac83} \S 1.7),
(i) and (ii) imply
\[ \mathbb{R} \pi_* \mathbb{Q}_{X(r)} 
  \simeq \bigoplus_{l=0, 2, 4, \cdots} 
  (j_{X_l})_{!*} \mathbb{Q}_{X_l}[-l].
\]
In view of (iii),
each direct summand $(j_{X_l})_{!*} \mathbb{Q}_{X_l}$
must be isomorphic to $\mathbb{Q}_{W_{r-l}}$
without any higher cohomology sheaf.

\subsection{Lemmas on $W_r$}
We introduce two auxiliary lemmas
concerning the cohomology of $W_r.$
Let $a_r^k: H^k(W_r) \to H^k(W_{r-1})$ 
and $b_r^k: H^k(J) \to H^k(W_r)$ 
be the maps induced by the inclusions 
$W_{r-1} \to W_r$ and $W_r \to J.$

\begin{lemma}
  \label{cohom-Wr}
If $0 \leq k \leq r,$
the map $b_r^k: H^k(J) \to H^k(W_r)$ 
is an isomorphism.
If $r \leq k \leq 2r,$
there exists a canonical isomorphism
$\check{b}_r^k: H^{k+2(g-r)}(J)(g-r) 
  \overset{\sim}{\longrightarrow} H^k(W_r).$
(Here $(g-r)$ indicates the Tate twist of the Hodge structure
\cite{Deligne}.)
In particular,
the dimension of $H^k(W_r)$ is
$\begin{pmatrix} 2g \\ k \end{pmatrix}$
or
$\begin{pmatrix} 2g \\ 2r-k \end{pmatrix}$
according to 
$0 \leq k \leq r$ or $r \leq k \leq 2r.$
\end{lemma}
{\it Proof.}
When $r=g,$ the assertion is trivial.
We show the assertion by the decreasing induction on $r$.
We assume the assertion for $r.$
Thanks to Proposition \ref{affine},
we have $H_c^k(W_r \setminus W_{r-1})=0$ if $k<r.$
By the long exact sequence
\[ \cdots \to H_c^k(W_r \setminus W_{r-1})
    \to H^k(W_r) \overset{a_r^k}{\to} H^k(W_{r-1})
    \to H_c^{k+1}(W_r \setminus W_{r-1})
    \to \cdots,
\]
we see that $a_r^k$
is an isomorphism if $k \leq r-2$
and an injection if $k=r-1.$
By the Poincar\'e duality assured by Proposition \ref{ih}, 
we obtain a map 
$\check{a}_r^k: H^k(W_{r-1}) \to H^{k+2}(W_r)(1)$
which is an isomorphism if $k \geq r$
and a surjection if $k=r-1.$
It remains to show the bijectivity of
$a_{r}^{r-1}$ and $\check{a}_{r}^{r-1}.$
To show this, we compare the dimensions.
By the inductive hypothesis we have
$h^{r-1}(W_r) = h^{r+1}(W_r) = 
\begin{pmatrix} 2g \\ r-1\end{pmatrix}.$
Since we have proved the lemma 
for $b_{r-1}^k$ and $\check{b}_{r-1}^k$ with $k \not= r-1,$
we can compute $h^{r-1}(W_{r-1})$ as
\[  h^{r-1}(W_{r-1})
  = (-1)^{r-1} \Bigr( 
               \chi(W_{r-1}) 
               - \sum_{k \neq r-1}(-1)^k h^k(W_{r-1})
               \Bigr)
    = \begin{pmatrix}
    2g \\ r-1
   \end{pmatrix}.
\]
Here we used Proposition \ref{chi-W_k}.
This completes the proof. ~~ $\square$

\begin{lemma}
  \label{lemma:maps-W_r}
  The map $a_r^k: H^k(W_r) \to H^k(W_{r-1})$ 
  is surjective for any $k$ and $r.$
  (When $k \leq r-1,$ this is an isomorphism
  as proved in the above lemma.) 
\end{lemma}
{\it Proof.}
We fix $k.$
We consider the following commutative diagram:
\[
\begin{matrix}
&H^k(J) & \overset{b_r^k}{\to}& H^k(W_r) 
    &\overset{a_r^k}{\to} & H^k(W_{r-1}) \\
&{}_c \downarrow & & \downarrow \\
&H^k(X(g)) &\overset{d_r}{\to}& H^k(X(r)).
\end{matrix}
\]
It is enough to show the surjectivity of 
$a_r^k \circ b_r^k = b_{r-1}^k$
for all $r.$
Thus we shall show the surjectivity of $b_r^k$ instead.
The diagram shows the inequality
\[ \dim \mathrm{Im} (d_r \circ c) 
   \leq \dim \mathrm{Im} b_r^k \leq h^k(W_r). 
\]
By Proposition \ref{lemmac} and Lemma \ref{cohom-Wr},
we see the dimension of $\mathrm{Im} (d_r \circ c)$
coincides with $h^k(W_r).$
Therefore the equality holds in the inequality above,
and the proof is done. ~~$\square$

\begin{remark}\label{naka-rem}
Assume $k \geq r.$ 
Via the isomorphisms $\check{b}_r^k$ in Lemma \ref{cohom-Wr},
the map $a_r^k$ in Lemma \ref{lemma:maps-W_r} can be rewritten as
(a Tate twist of)
\[ H^{k+2(g-r)}(J) \to H^{k+2(g-r)+2}(J)(1). \]
This map seems to coincide with
the cup product with a hyperplane section,
up to a multiplication by a non-zero constant.
(This would imply Lemma \ref{lemma:maps-W_r} 
by the Hard Lefschetz Theorem.)
When $r=g,$ this was shown by Nakayashiki \cite{Nakayashiki00}.
\end{remark}

\subsection{Proof of \eqref{dimcoh1}}
We introduce the following notations:
\[ H_r^k = H^k(W_r), \qquad
   K_r^k = \ker[ H_r^k \overset{a_r^k}{\twoheadrightarrow} H_{r-1}^k].
\]
(If $k \leq r-1$ we have $K_r^k = 0.$)
Recall that $a_r^k$ is surjective by Lemma \ref{lemma:maps-W_r}.
By Lemma \ref{cohom-Wr} we see the dimension of $K_r^k$
is zero if $0 \leq k \leq r-1,$
and is
$\begin{pmatrix} 2g \\ 2r-k \end{pmatrix}
-\begin{pmatrix} 2g \\ 2r-k-2 \end{pmatrix}$
if $r \leq k \leq 2r.$
The long exact sequence
\[ \cdots \to H_c^k(W_r \setminus W_{r-1})
   \to H^k(W_r) \overset{a_r^k}{\to} H^k(W_{r-1}) 
   \to H_c^{k+1}(W_r \setminus W_{r-1}) \to \cdots
\]
provides an isomorphism
\[  H_c^k(W_r \setminus W_{r-1}) \cong K_r^k. \]
By the Poincar\'e duality (for the usual cohomology theory),
we see
$h^k(W_r \setminus W_{r-1}) = h_c^{2r-k}(W_r \setminus W_{r-1}).$
This proves \eqref{dimcoh1}.
We remark that our proof in the case of $r=g$ is
basically same as Nakayashiki's proof,
except that the use of the Hard Lefschetz theorem was
avoided in Lemma \ref{lemma:maps-W_r}
(see Remark \ref{naka-rem}).

\subsection{Proof of \eqref{dimcoh2}}
According to \eqref{thetalocation},
we have an exact sequence of sheaves on $W_r^{\pm}:$
\[
    0 \longrightarrow \mathbb{Q}_{W_r^{\pm}}
    \longrightarrow \mathbb{Q}_{W_r^+}
       \oplus \mathbb{Q}_{W_r^-}
    \longrightarrow \mathbb{Q}_{W_{r-1}} \to 0.
\]
(For simplicity we write $\mathbb{Q}_S$ 
instead of $\iota_* \mathbb{Q}_S$
for a closed immersion $\iota: S \hookrightarrow T.$)
The long exact sequence deduced from this implies
(again by Lemma \ref{lemma:maps-W_r})
\[ H^k(W_r^{\pm}) 
   \cong H_r^k \oplus K_r^k. 
\]
We then consider the long exact sequence
\[ \cdots \to H_c^k(W_r \setminus W_{r-1}^{\pm})
   \to H^k(W_r)
   \overset{f_r^k}{\to} H^k(W_{r-1}^{\pm}) 
   \to H_c^{k+1}(W_r \setminus W_{r-1}^{\pm})
   \to \cdots.
\]
With respect to the isomorphisms
$H^k(W_r) \cong H_r^k$ and 
$H^k(W_{r-1}^{\pm}) \cong H_{r-1}^{k} \oplus K_{r-1}^k,$
the map $f_r^k$ reads as 
$f_r^k(x) = (a_r^k(x), 0).$
Therefore we obtain an exact sequence
\[ 0 \to K_{r-1}^{k-1} 
     \to H_c^k(W_r \setminus W_{r-1}^{\pm})
     \to K_r^k \to 0.
\]
In particular, 
$h_c^k(W_r \setminus W_{r-1}^{\pm})$
is
$\begin{pmatrix} 2g+1 \\ 2r-k \end{pmatrix}
   -\begin{pmatrix} 2g+1 \\ 2r-k-2 \end{pmatrix}$
if $r \leq k \leq 2r,$
and is zero otherwise.
The Poincar\'e duality shows
$h^k(W_r \setminus W_{r-1}^{\pm})
=h_c^{2r-k}(W_r \setminus W_{r-1}^{\pm}).$
This proves \eqref{dimcoh2}.

\subsection{Proof of \eqref{dimcoh3}}
The first step is to study the cohomology of $W_r^{0+}.$
The relation \eqref{thetalocation} gives 
a resolution of the sheaf $\mathbb{Q}_{W_r^{0+}}:$
\[ 0 \to \mathbb{Q}_{W_r^{0+}}
  \to \mathbb{Q}_{W_r} \oplus \mathbb{Q}_{W_r^+} 
  \to \mathbb{Q}_{W_{r-1}} \oplus \mathbb{Q}_{W_{r-1}^+} 
  \to \mathbb{Q}_{W_{r-2}} \oplus \mathbb{Q}_{W_{r-2}^+} 
  \to \cdots.
\]
We consider the deduced spectral sequence 
\[ E_1^{i,j} = H^j(W_{r-i}) \oplus H^j(W_{r-i}^{+})
   \Rightarrow H^{i+j}(W_r^{0+}).
\]
If we identify $E_1^{i,j}$ with $H_{r-i}^{j \oplus 2},$
the boundary map $d_1^{i,j}$ reads as
\[ H_{r-i}^{j \oplus 2} \to H_{r-i-1}^{j \oplus 2}
   ~;~
   (x,y) \mapsto (a_{r-i}^j(x-y), a_{r-i}^j(x-y)).
\]
Therefore we obtain
\[ E_2^{i,j} = 
\left\{
\begin{array}{c}
H_r^j \oplus K_r^j \qquad (i=0) \\
K_{r-i}^j     ~~~~~\qquad (i>0)
\end{array}
\right. 
\]
Since the weight of the Hodge structure on $E_2^{i,j}$
are different for different $j$'s
by Proposition \ref{ih},
we have the degeneration of the spectral sequence at $E_2$-terms
and the decomposition
\cite{Deligne}

\[ H^k(W_r^{0+})  
  \cong H_r^k \oplus \bigoplus_{i=0}^{\min(k, r)} K_{r-i}^{k-i}.
\]
The same formula holds also for $H^k(W_r^{0-}).$

The rest of the proof is similar to the previous subsection.
By \eqref{thetalocation},
we have a long exact sequence
\[ \cdots \to H^k(W_r^{0 \pm}) 
    \to H^k(W_r^{0+}) \oplus H^k(W_r^{-})
    \to H^k(W_{r-1}^{0-})
    \to H^{k+1}(W_r^{0 \pm}) 
    \to \cdots,
\]
and hence an exact sequence
\[ 0 \to \bigoplus_{i=1}^{min(k, r)} K_{r-i}^{k-i}
     \to H^k(W_r^{0 \pm}) 
     \to H_r^k \oplus K_r^{k \oplus 2}
         \oplus \bigoplus_{i=1}^{min(k, r)} K_{r-i}^{k-i}
     \to 0.
\]
We put this description in the long exact sequence
\[ \cdots \to H_c^k(W_r \setminus W_{r-1}^{0 \pm})
   \to H^k(W_r)
   \overset{g_r^k}{\to} H^k(W_{r-1}^{0 \pm}) 
   \to H_c^{k+1}(W_r \setminus W_{r-1}^{0 \pm})
   \to \cdots.
\]
Then we get
\[
\dim \ker g_r^k = \dim K_r^k, 
\qquad
\dim \coker g_r^k = 2 \sum_{i=0}^{\min(k, r)} \dim K_{r-i-1}^{k-i},
\]
so
$h_c^k(W_r \setminus W_{r-1}^{0 \pm})
  = \dim \ker g_r^k + \dim \coker g_r^{k-1}$
can be determined
by a straight forward computation.
Now the Poincar\'e duality 
$h^k(W_r \setminus W_{r-1}^{0 \pm})
  = h_c^{2r-k}(W_r \setminus W_{r-1}^{0 \pm})$
completes the proof of \eqref{dimcoh3}.
~~$\square$


\section{Mumford system and its variants}

\subsection{Mumford system}
The Mumford system \cite{Mumford-Book}
is described by the Lax matrix \eqref{Lax-Mum} with
\begin{align}
  \label{poly-oddMum}
  \begin{split}
  &u(x) = x^g + u_1 x^{g-1} + \cdots + u_{g} 
  \\
  &v(x) = v_{\frac{3}{2}} x^{g-1} + v_{\frac{5}{2}} x^{g-2} 
          + \cdots + v_{g+\frac{1}{2}}
  \\
  &w(x) = x^{g+1} + w_1 x^{g} + \cdots + w_{g+1}.
  \end{split}
\end{align}
The coefficients of these polynomials constitute the phase space
$\mathcal{M}_g \simeq \mathbb{C}^{3g+1}$.
We also consider the space
of monic polynomials
$\mathbb{C}[x]_{\deg = 2g+1}^{\text{monic}} 
\cong \mathbb{C}^{2g+1}$
of degree $2g+1.$
We define a map
\begin{align}
  \label{Mum-phi}
  \psi : \mathcal{M}_g 
    \to \mathbb{C}[x]_{\deg = 2g+1}^{\text{monic}}
  ~;~ 
  l(x) \mapsto u(x) w(x) + v(x)^2.
\end{align}
Recall \eqref{Mum-levelset} that 
for $f(x) = x^{2g+1} + f_1 x^{2g} + \cdots + f_{2g+1} 
\in \mathbb{C}[x]_{\deg = 2g+1}^{\text{monic}},$
the fiber $\psi^{-1}(f)$ 
is denoted by $\mathcal{M}_{g,f}.$
Assume $f(x)$ has no multiple zero,
and let $X$ be the hyperelliptic curve $X$ of genus $g$
defined by $y^2 = f(x).$
Let $J(X)$ and $\Theta$ be the Jacobi variety of $X$
and its theta divisor. 
As mentioned in the introduction, we have
the following theorem.

\begin{theorem}\cite{Mumford-Book}
\label{oddMum-Jac}
$\mathcal{M}_{g,f}$ is isomorphic to $J(X) \setminus \Theta.$
\end{theorem}

The space $\mathcal{M}_g$ is 
equipped with the structure of a dynamical system
by the following theorem.
\begin{theorem}(\cite{Mumford-Book} Theorem 3.1)
\label{vectorfield}
  There are the independent and commuting
  invariant vector fields $D_1,\ldots,D_g$
  on $\mathcal{M}_g$ given by
  \begin{align}
  \label{time-evol}
  \begin{split}
    &D(x_2) u(x_1) = \frac{u(x_1) v(x_2) - v(x_1) u(x_2)}{x_1-x_2}
    \\
    &D(x_2) v(x_1) = \frac{1}{2} \Bigl(
                 \frac{w(x_1) u(x_2) - u(x_1) w(x_2)}{x_1-x_2} 
                 - \alpha(x_1+x_2) u(x_1) u(x_2)
                 \Bigr)
    \\
    &D(x_2) w(x_1) = \frac{v(x_1) w(x_2) - w(x_1) v(x_2)}{x_1-x_2} 
                     + \alpha(x_1+x_2) v(x_1) u(x_2)
  \end{split}
  \end{align}
  where $D(x) = \sum_{i=1}^g x^{g-i} D_i$, and $\alpha(x) = 1$.
\end{theorem}
To make the notion of the dynamical system clearer,
we introduce $g$ times $t_1, \ldots t_g$ given by   
$\frac{\partial \mathcal{o}}{\partial t_i} = D_i \mathcal{o}$.
The coefficients of $u(x) w(x) + v(x)^2$
are the invariants of $D_i.$
Thus the vector fields are well-defined on 
the level set $\mathcal{M}_{g,f}$.
It is shown in \cite{Mumford-Book} \S 5 that 
if $f(x)$ has no multiple zero,
then the flow of $D_i$ is linearized on $J(X) \setminus \Theta$.

The map $\psi: \mathcal{M}_g \to 
\mathbb{C}[x]_{\deg = 2g+1}^{\text{monic}}$
induces an inclusion between their affine rings
$$
\mathcal{A} = 
\mathbb{C}[u_1,\ldots,u_g,v_{\frac{3}{2}},\ldots,v_{g+\frac{1}{2}},
w_1,\ldots,w_{g+1}]
\hookleftarrow
\mathcal{F} = \mathbb{C}[f_1, \ldots, f_{2g+1}].
$$
The actions of $D_i$ given by \eqref{time-evol} 
are naturally extended to $\mathcal{A}$.
Let $C^1$ be the free $\mathcal{A}$-module
with the basis $d t_1, \ldots, d t_g,$
and let $C^k = \wedge^k C^1.$
We define the complex
\begin{align}
  \label{complex-A}
  0  \longrightarrow C^0 \stackrel{d}{\longrightarrow} C^1  
            \stackrel{d}{\longrightarrow} \cdots
            \stackrel{d}{\longrightarrow} C^{g-1} 
            \stackrel{d}{\longrightarrow} C^g \longrightarrow 0,
\end{align}
where the differential $d$ is given by
\begin{align}
  \label{d-operator}
  d = \sum_{i=1}^g d t_i \wedge D_i ~:~ C^k \to C^{k+1}.
\end{align}

Let $f(x)$ be a monic polynomial of degree $2g+1.$
We write $\mathcal{A}_f$ for the affine ring of $\mathcal{M}_{g, f}.$
Then $\mathcal{A}_f$ is a quotient ring of 
$\mathcal{A}$ divided by the relation
$u(x) w(x) + v(x)^2 = f(x).$
By tensoring $\mathcal{A}_f$ 
over $\mathcal{A}$ with \eqref{complex-A},
we get the complex
\[
  0  \longrightarrow C^0_f \stackrel{d}{\longrightarrow} C^1_f  
            \stackrel{d}{\longrightarrow} \cdots
            \stackrel{d}{\longrightarrow} C^{g-1}_f 
            \stackrel{d}{\longrightarrow} C^g_f \longrightarrow 0.
\]
We note that the highest cohomology group of this complex
is isomorphic to the space mentioned in the introduction
\[ H^g(C_f^*) 
  \simeq \mathcal{A}_f / \sum_{i=1}^g D_i \mathcal{A}_f.
\]
Furthermore, when $f(x)$ has no multiple zero
we have an isomorphism 
\begin{align*}
  H^k(J(X) \setminus \Theta, \mathbb{C}) \simeq H^k(C_f^\ast) 
\end{align*}
due to Theorem \ref{oddMum-Jac}
and the algebraic de Rham theorem.
This cohomology was computed in Theorem \ref{theorem:oddMum-H^k}.

We define a grading on $\mathcal{A}$ by setting
$\deg(*_i) = i ~$ for $* \in \{u, v, w\}.$
Then $\mathcal{F} \subset \mathcal{A}$
is a graded subring and $\deg(f_i)=i.$
We set $\deg(D_i) = i-\frac{1}{2}$
(note that 
$D_i \mathcal{A}^{(j)} \subset \mathcal{A}^{(i+j-\frac{1}{2})}$),
and $\deg({d t_i}) = -i + \frac{1}{2}$ 
so that the degree of $d$ is zero.
We write $\mathcal{A}_0$ for 
$\mathcal{A}_{x^{2g+1}} 
= \mathcal{A} / \sum_{i=1}^{2g+1} f_i \mathcal{A},$
which becomes a graded ring.
Similarly, we write $C_0^*$ for the complex $C_{x^{2g+1}}^*.$
The $q$-Euler characteristic of $C_0^*$
is defined to be
$\chi_q (C_0^\ast) = \sum_{k=0}^g (-1)^k \ch(C_0^k)$,
where we write
$\ch(\mathcal{O}) = \sum_{k=0}^{\infty} q^k \dim \mathcal{O}^{(k)}$
for a graded space $\mathcal{O} = \oplus_{k \geq 0} \mathcal{O}^{(k)}.$
The notations
\begin{align*}
  &[k]_q = (1-q^k),
  \\
  &[k]_q! = 
  \begin{cases}
  [k]_q [k-1]_q \cdots [1]_q ~~ \text{for } k \in \mathbb{Z}_{>0},
  \\[0mm]
  [k]_q [k-1]_q \cdots [\frac{1}{2}]_q 
  ~~ \text{for } k \in \frac{1}{2} + \mathbb{Z}_{\geq 0},
  \end{cases}
\end{align*}
are used to state the following result of
Nakayashiki-Smirnov.
\begin{theorem}
(\cite{SmirnovNakayashiki00} eq. (17,18).)
  \label{Euler-C0}
\begin{align*}
  \chi_q(C_0^\ast) = (-1)^g q^{-\frac{1}{2}g^2}
                    \frac{[\frac{1}{2}]_q[2g+1]_q !}
                         {[g+\frac{1}{2}]_q! [g]_q![g+1]_q!}
                  ~\stackrel{q \to 1} \longrightarrow~ 
                 (-1)^g 
                 \Bigl(
                 \begin{pmatrix}
                   2g \\ g
                 \end{pmatrix}
                 -
                 \begin{pmatrix}
                   2g \\ g-1
                 \end{pmatrix}
                \Bigr).
\end{align*}
\end{theorem}
Note that this limit coincides with 
$\chi(J(X)\setminus \Theta)$ 
in Theorem \ref{theorem:oddMum-H^k}.
This suggests 
$\dim H^k(C_0^\ast) = \dim H^k(C_f^\ast),$
although we do not even have the finiteness of $\dim H^k(C_0^\ast).$

\subsection{Even Mumford system}

The even Mumford system was first introduced in 
\cite{Van92} when $g=2,$
and generalized to general $g$ in \cite{Vanhae01}.
This system is described by the Lax matrix 
$l(x)$ \eqref{Lax-Mum} where the polynomials 
$u(x), v(x)$ and $w(x)$ are set to be
\begin{align*}
  \begin{split}
  &u(x) = x^g + u_1 x^{g-1} + \cdots + u_{g} 
  \\
  &v(x) = v_{2} x^{g-1} + v_{3} x^{g-2} + \cdots + v_{g+1}
  \\
  &w(x) = x^{g+2} + w_1 x^{g+1} + \cdots + w_{g+2}.
  \end{split}
\end{align*}
The coefficients of these polynomials constitute the phase space
$\mathcal{M}_g^\prime \simeq \mathbb{C}^{3g+2}$.
We consider the space 
$\mathbb{C}[x]_{\deg = 2g+2}^{\text{monic}}$
of monic polynomials of degree $2g+2.$
The fiber of the map
$$
  \mathcal{M}_g^\prime 
      \to \mathbb{C}[x]_{\deg = 2g+2}^{\text{monic}} 
  ~;~
         l(x) \mapsto u(x) w(x) + v(x)^2
$$
over
$f(x)  = x^{2g+2} + f_1 x^{2g+1} + \cdots + f_{2g+2}
\in \mathbb{C}[x]_{\deg = 2g+2}^{\text{monic}}$
is denoted by $\mathcal{M}_{g,f}^\prime.$
Assume $f(x)$ has no multiple zero,
and let $X$ be the hyperelliptic curve of genus $g$
defined by $y^2 = f(x).$
Note that there are two points $\infty_+, \infty_- \in X$ above 
$\infty \in \mathbb{P}^1.$
Let $J(X)$ be the Jacobi variety of $X,$
and let $\Theta$ be the theta divisor.
We also write $\Theta^\prime$
for the divisor defined in Theorem \ref{theorem:evenMum-H^k}.
The affine Jacobi variety $J(X) \setminus \Theta'$ 
has a matrix realization given by $\mathcal{M}_{g, f}'$:
\begin{theorem}(\cite{Vanhae01} Proposition 3.1)
  \label{theorem:evenMum}
  $\mathcal{M}_{g,f}^\prime$ 
  is isomorphic to $J(X) \setminus \Theta^\prime$.
\end{theorem}

Parallel to the Mumford system, we have 
$g$ commuting invariant vector fields $D_1,\ldots,D_g$
on $\mathcal{M}_g^\prime$ \cite{Vanhae01}.
Their action is written as \eqref{time-evol} 
by using the operator $D(x) = \sum_{i=1}^g x^{g-i} D_i$ and
$\alpha(x) = x + w_1 - u_1$.  
 
We write the affine rings of $\mathcal{M}_g'$
and $\mathbb{C}[x]_{\deg = 2g+2}^{\text{monic}}$ as
$$
 \mathcal{A}^\prime = 
 \mathbb{C}[u_1,\ldots,u_g,v_2,\ldots,v_{g+1},w_1,\ldots,w_{g+2}],
 ~~
 \mathcal{F}^\prime = \mathbb{C}[f_1,\ldots,f_{2g+2}].
$$
We introduce a grading on $\mathcal{A}^\prime$ and $\mathcal{F}'$
by setting $\deg(*_i)=i$ for $* \in \{u, v, w, f\},$
which is compatible with 
the inclusion $\mathcal{F}' \subset \mathcal{A}'$.
Based on \eqref{time-evol},
we set $\deg(D_i)=i.$
We define $\mathcal{A}_f^\prime$ 
to be the affine ring of $\mathcal{M}_{g, f}^\prime$,
and let $\mathcal{A}_0^\prime = \mathcal{A}_{x^{2g+2}}^\prime.$
We also define complexes $C_f^{*'}$ and $C_0^{*'}$
in the same manner as the previous subsection.
Then we have an isomorphism
\[ H^g(C_f^{*'}) 
   \cong \mathcal{A}_f'/ \sum_{i=1}^g D_i \mathcal{A}_f', 
\]
which is a motivation for the study of $H^k(C_f^{*'}).$
When $f(x)$ has no multiple zero,
Theorem \ref{theorem:evenMum} gives an isomorphism 
\[ H^k(J(X) \setminus \Theta', \mathbb{C}) \cong H^k(C_f^{*'}), \]
which was computed in Theorem \ref{theorem:evenMum-H^k}.
On the other hand,
the $q$-Euler characteristic of $C_0^{*'}$
can be computed by using the method of 
\cite{SmirnovNakayashiki02} eq. (3.2).
\begin{theorem}
(\cite{Nakayashiki-private})
  \label{evenMum-Euler}
\begin{align*}
  \chi_q(C_0^{*'}) = (-1)^g q^{-\frac{1}{2}g(g+1)}
                    \frac{[1]_q[2g+2]_q !}{[g+1]_q! [g+2]_q!}
                  ~\stackrel{q \to 1} \longrightarrow~ 
                  (-1)^g
                  \Bigl( 
                  \begin{pmatrix}
                   2g+1 \\ g
                 \end{pmatrix}
                 -
                 \begin{pmatrix}
                   2g+1 \\ g-1
                 \end{pmatrix}
                 \Bigr).
\end{align*}
\end{theorem}
Again this limit coincides with 
$\chi(J(X) \setminus \Theta^\prime)$
in Theorem \ref{theorem:evenMum-H^k}.

\subsection{Hyperelliptic Prym systems}

We study two types of hyperelliptic Prym systems 
introduced in \cite{Vanhae01}.
These systems are obtained as subsystems  of the even Mumford system
$\mathcal{M}_n^\prime.$
We consider two cases,
(I)  the case of $n=2g$ and (II)  the case of $n=2g+1$.
The Lax matrices for the hyperelliptic Prym systems
are written as \eqref{Lax-Mum} with the polynomials 
$u(x), v(x)$ and $w(x)$ given by
\begin{align*}
  &\begin{cases}
  u(x) = x^{2g} + u_1 x^{2g-2} + u_2 x^{2g-4} + \cdots + u_{g} 
  \\
  v(x) = v_{1} x^{2g-1} + v_{2} x^{2g-3} + \cdots + v_{g} x
  \\
  w(x) = x^{2g+2} + w_1 x^{2g} + w_2 x^{2g-2} + \cdots + w_{g+1}
  \end{cases}
  ~~\text{ for (I)},
  \\
  &\begin{cases}
  u(x) = x^{2g+1} + u_1 x^{2g-1} + u_2 x^{2g-3} + \cdots + u_{g} x 
  \\
  v(x) = v_{1} x^{2g} + v_{2} x^{2g-2} + \cdots + v_{g+1}
  \\
  w(x) = x^{2g+3} + w_1 x^{2g+1} + w_2 x^{2g-1} + \cdots + w_{g+1} x
  \end{cases}
  ~~\text{ for (II)}.
\end{align*}
The coefficients of these polynomials respectively 
constitute affine spaces 
$\mathcal{P}_g^{(I)} \simeq \mathbb{C}^{3g+1} 
\subset \mathcal{M}_{2g}^\prime$
and 
$\mathcal{P}_g^{(II)} \simeq \mathbb{C}^{3g+2} 
\subset \mathcal{M}_{2g+1}^\prime$.
Let $f(x) \in \mathbb{C}[x^2]$ be of the form
\begin{align}
  \label{Prym-f}
  f(x) = 
  \begin{cases}
  x^{4g+2} + f_{1} x^{4g} + f_{2} x^{4g-2} + \cdots + f_{2g+1}
  ~~ \text{for (I)},
  \\
  x^{4g+4} + f_{1} x^{4g+2} + f_{2} x^{4g} + \cdots + f_{2g+2}
  ~~ \text{for (II)}.      
  \end{cases}
\end{align}
We assume $f(x)$ has no multiple zero,
and let $X$ be the hyperelliptic curve defined by $y^2 = f(x).$
There are two points $\infty_1, \infty_2 \in X$
above $\infty \in \mathbb{P}^1,$
and two points $O_1, O_2 \in X$ above
the point $x=0$ on $\mathbb{P}^1.$
There are two involutions $\sigma$ and $\tau$ on $X$
other than the hyperelliptic involution $\iota$:
\begin{center}
\begin{tabular}{c|c|c}
  & (I) $n=2g$ & (II) $n=2g+1$ \\
  \hline
  $\iota : (x,y) \mapsto$ & $(x,-y)$ & $(x,-y)$ \\[1mm]
  $\sigma : (x,y) \mapsto$ & $(-x,y)$ & $(-x,-y)$ \\[1mm]
  $\tau : (x,y) \mapsto$ & $(-x,-y)$ & $(-x,y)$
\end{tabular}
\end{center}
We write 
$\pi_{\sigma}: X \to X_{\sigma} = X/\sigma$
and
$\pi_{\tau}: X \to X_{\tau} = X/\tau$
for the quotient maps.
Note that $X_{\tau}$ is a hyperelliptic curve of genus $g,$
that $O=\pi_{\tau}(O_1)=\pi_{\tau}(O_2)$
is a Weierstrass point, and that 
$\infty_+ = \pi_{\tau}(\infty_1)$
is conjugate to $\infty_- = \pi_{\tau}(\infty_1)$
under the hyperelliptic involution on $X_\tau.$

We recall that the hyperelliptic Prym variety 
$\Prym(X/X_\sigma)$ of $(X, \sigma)$ 
is a sub abelian variety of $J(X)$
defined as
$$
    \Prym(X/X_\sigma) = \{ D - \sigma(D) | D \in J(X) \}.
$$
The following result was first obtained by Mumford for 
the (I) case, and by Dalaljan for the (II) case.
See \cite{Vanhae01} for an unified treatment.
\begin{theorem}(\cite{Vanhae01} Theorem 2.5.)
  \label{theorem:PrymJac-iso}
  The map $\pi_{\tau}$ induces an isomorphism
  $$
    \pi_{\tau}^* : ~J(X_\tau) \to \Prym(X/X_\sigma)
    \qquad
          [p] \mapsto [\pi_\tau^\ast(p)].
  $$
  Moreover, $\pi_{\tau}^*$ maps
  $J(X_\tau) \setminus \Theta^{\prime\prime}$
  isomorphically onto
  $\Prym(X/X_{\sigma}) \setminus \Theta^{\prime},$
  where $\Theta^{\prime\prime}$
  is a divisor defined in Theorem \ref{theorem:Prym-H^k}.
\end{theorem}
For $i= I$ and  $II$,
the level set 
$\mathcal{P}_{g,f}^{(i)}
= \{ l(x) \in \mathcal{P}_{g}^{(i)}~|~ u(x) w(x) + v(x)^2 = f(x) \}$
gives the matrix realization 
of the Prym variety:
\begin{theorem} (\cite{Vanhae01} Propositions 3.2, 3.3)
  \label{theorem:prym-iso}
  $\mathcal{P}_{g,f}^{(I)}$
  is isomorphic to
  $\Prym(X/X_{\sigma}) \setminus \Theta^{\prime},$
  and
  $\mathcal{P}_{g,f}^{(II)}$ is isomorphic to
  a disjoint union of two translates of
  $\Prym(X/X_{\sigma}) \setminus \Theta^{\prime}.$
\end{theorem}

The invariant vector fields on $\mathcal{P}_g^{(I)}$ or
$\mathcal{P}_g^{(II)}$ 
are respectively obtained by reducing those on $\mathcal{M}_{n}^\prime$.
The action of the commuting operators $D_1,\ldots,D_g$ is written 
in a common way in both cases as \cite{Vanhae01}:
\begin{align}
  \label{time-evol-prym}
  \begin{split}
    &D(x_2) u(x_1) = \frac{x_2 u(x_1) v(x_2) - x_1 v(x_1) u(x_2)}
                          {x_1^2-x_2^2}
    \\
    &D(x_2) v(x_1) = \frac{x_1}{2} \Bigl(
                 \frac{w(x_1) u(x_2) - u(x_1) w(x_2)}{x_1^2-x_2^2} 
                 - u(x_1) u(x_2)
                 \Bigr)
    \\
    &D(x_2) w(x_1) = \frac{x_1 v(x_1) w(x_2) - x_2 w(x_1) v(x_2)}{x_1^2-x_2^2} 
                     + x_1 v(x_1) u(x_2),
  \end{split}
\end{align}
where 
\begin{align*}
  D(x) = 
  \begin{cases}
  \displaystyle{\sum_{i=1}^g x^{2(g-i)}} D_i ~~~ \text{for (I)},
  \\
  \displaystyle{\sum_{i=1}^g x^{2(g-i)+1}} D_i ~~~ \text{for (II)}.
  \end{cases}
\end{align*}

Starting with the rings
\begin{align*}
  &\mathcal{A}^{(I)} =
  \mathbb{C}[u_1,\ldots,u_g,v_1,\ldots,v_{g},w_1,\ldots,w_{g+1}],
  ~~
 \mathcal{F}^{(I)} = \mathbb{C}[f_1,\ldots,f_{2g+1}],
  \\
 &\mathcal{A}^{(II)} =
  \mathbb{C}[u_1,\ldots,u_g,v_1,\ldots,v_{g+1},w_1,\ldots,w_{g+1}],
  ~~
 \mathcal{F}^{(II)} = \mathbb{C}[f_1,\ldots,f_{2g+2}],
\end{align*}
we define the complexes $C^{*''}, C^{*''}_{f}, C^{*''}_0$
by a similar method as the previous subsections.
When $f(x)$ has no multiple zero,
Theorem \ref{theorem:prym-iso}
shows that $H^k(C_f^{*''})$
is isomorphic to
$H^k(J(X_{\tau}) \setminus \Theta'', \mathbb{C})$ for (I),
and to a direct sum of two copies of
$H^k(J(X_{\tau}) \setminus \Theta'', \mathbb{C})$ for (II).
Their dimension is calculated in Theorem \ref{theorem:Prym-H^k}.

We define the grading on 
$\mathcal{A}^{(I)}, \mathcal{A}^{(II)},
 \mathcal{F}^{(I)}$ and $\mathcal{F}^{(II)}$
by setting $\deg(*_i) = i$ for $* \in \{u, v, w, f\}.$
We also define $\deg(D_i)=i$
based on \eqref{time-evol-prym}.
By using \cite{SmirnovNakayashiki02} eq. (3.2),
we can show the following
\begin{theorem}
\label{Prym-Euler}
\begin{align*}
    \chi_q(C_0^{*''}) =
                     \begin{cases}
                      (-1)^g q^{-\frac{1}{2} g(g+1)}
                       \displaystyle{\frac{[2g+1]_q!}{[g]_q![g+1]_q!}}
                     ~\stackrel{q \to 1} \longrightarrow~ 
                     (-1)^g
                     \begin{pmatrix}
                       2g+1 \\ g
                     \end{pmatrix}
                     ~~ \text{for (I)}, 
                     \\
                      (-1)^g q^{-\frac{1}{2} g(g+1)}
                       \displaystyle{\frac{[2g+2]_q!}{[g+1]_q![g+1]_q!}}
                     ~\stackrel{q \to 1} \longrightarrow~ 
                     (-1)^g
                     \begin{pmatrix}
                       2g+2 \\ g+1
                     \end{pmatrix}
                     ~~ \text{for (II)}. 
                     \end{cases}
\end{align*}
\end{theorem}
Theorems \ref{theorem:Prym-H^k} and \ref{theorem:prym-iso}
show that this limit coincides with 
the Euler characteristic of the generic level set 
in both cases.


\section{Noumi-Yamada system at $\alpha = 0$}

\subsection{Mumford system and the systems 
$\mathcal{D}_g^{(I)}, \mathcal{D}_g^{(II)}$}

We introduce integrable systems $\mathcal{D}_g^{(I)}$ and  
$\mathcal{D}_g^{(II)}$
motivated by the study \cite{Takasaki03} on 
the Noumi-Yamada systems.
We relate these systems to
the Mumford system $\mathcal{M}_g$.
This relation is applied in the next two subsections
to study the integrability and the solution  
of the systems $\mathcal{Q}_g^{(I)}$ and $\mathcal{Q}_g^{(II)}$ 
which are the $\alpha = 0$ limit of 
the Noumi Yamada systems \eqref{NY-I} and \eqref{NY-II}.

We define the system  $\mathcal{D}_g^{(i)}$ for $i=I,II$
with the Lax matrix 
\begin{align}
  T(x) =
  \begin{pmatrix}
    a(x) & b(x) \\
    c(x) & d(x)
  \end{pmatrix}
  \in M_2(\mathbb{C}[x]),
\end{align}
where 
\begin{align*}
  \begin{split}
  &\begin{cases}
  &a(x)
  =
  a_{\frac{1}{2}} x^g + a_{\frac{3}{2}} x^{g-1} + \cdots + a_{\frac{2g+1}{2}},
  ~~
  b(x)
  = x^g + b_1 x^{g-1} + \cdots + b_g,
  \\
  &c(x)
  =
  x^{g+1} + c_1 x^g + \cdots + c_{g+1},
  ~~
  d(x)
  =
  a_{\frac{1}{2}} x^g + d_{\frac{3}{2}} x^{g-1} + \cdots + d_{\frac{2g+1}{2}},
  \end{cases}
  \text{for (I)},
  \\
  &\begin{cases}
  &a(x)
  =
  x^{g+1} + a_{\frac{1}{2}} x^g + a_{\frac{3}{2}} x^{g-1} 
  + \cdots + a_{\frac{2g+1}{2}},
  ~~
  b(x)
  = b_0 x^g + b_1 x^{g-1} + \cdots + b_g,
  \\
  &c(x)
  =
  b_0 x^{g+1} + c_1 x^g + \cdots + c_{g+1},
  ~~
  d(x)
  =
  x^{g+1} + a_{\frac{1}{2}} x^g + d_{\frac{3}{2}} x^{g-1} 
  + \cdots + d_{\frac{2g+1}{2}},
  \end{cases}
  \text{for (II)}.
  \end{split}
\end{align*}
The coefficients of the matrix entries constitute the 
phase space $\mathcal{D}_g^{(I)} \simeq \mathbb{C}^{4g+2}$ and 
$\mathcal{D}_g^{(II)} \simeq \mathbb{C}^{4g+3}$.
\begin{proposition}
  \label{prop:Dg-flow}
  On the space $\mathcal{D}_g^{(i)}$ for $i=I,II$,
  there are commuting and invariant vector fields
  $D_1, \cdots, D_g$ defined as 
  \begin{align}
  \label{Dg-flow}
    \begin{split}
  &D(x_2) a(x_1) = 
    \frac{c(x_1) b(x_2) - b(x_1) c(x_2)}{2(x_1-x_2)} - \frac{1}{2}b(x_1) b(x_2)
  \\
  &D(x_2) b(x_1) 
  = \frac{b(x_1) \bigl(a(x_2)-d(x_2)\bigr) 
          - \bigl(a(x_1) - d(x_1)\bigr) b(x_2)}{2(x_1-x_2)}
  \\
  &D(x_2) c(x_1) 
  = \frac{\bigl(a(x_1)-d(x_1)\bigr) c(x_2) 
          - c(x_1) \bigl(a(x_2)-d(x_2)\bigr)}{2(x_1-x_2)} 
         + \frac{1}{2}\bigl(a(x_1)-d(x_1)\bigr) b(x_2)
  \\
  &D(x_2) d(x_1) 
  = \frac{b(x_1) c(x_2) - c(x_1) b(x_2)}{2(x_1-x_2)} + \frac{1}{2}b(x_1) b(x_2)
    \end{split}
  \end{align}
  where $D(x) = \sum_{i=1}^g x^{g-i} D_i$.
  Especially, in the (II) case $b_0$ is a constant
  of these vector fields. 
\end{proposition}
{\it Proof.}
The commutativity of $D_i$'s
is proved by $[D(x_1), D(x_2)] = 0$ using \eqref{Dg-flow}.
Eqs. \eqref{Dg-flow} also indicate that
the spectral curve of $T(x)$ is invariant under 
the vector fields $D_1,\ldots, D_g$.
Though $b_0$ is not a coefficient of the spectral curve,
we see its invariance under the vector field from the second equation
in \eqref{Dg-flow}. 
~~ $\square$

Now we relate $\mathcal{D}_g^{(I)}$ to
the Mumford system $\mathcal{M}_g$ considered in \S 3.1.
The relation is summarized by the commutative diagram
\begin{equation}
\label{diagI}
\begin{matrix}
\mathcal{D}_g^{(I)} & \overset{\Phi}{\longrightarrow} & \mathcal{M}_g 
\\[1mm]
{}_{\psi'} \downarrow \quad & & ~ \downarrow_{\psi}
\\[1mm]
\mathcal{C}^{(I)}& 
\overset{\phi}{\longrightarrow} & 
~~~ \mathbb{C}[x]_{\deg = 2g+1}^{\text{monic}}.
\end{matrix}
\end{equation}
We explain each term in the diagram.
For each $n$, we set 
$\mathbb{C}[x]_{\deg \leq n}$ 
to be the space of polynomials in $x$ of degree $\leq n$.
We define $\mathcal{C}^{(I)} = \mathbb{C}[x]_{\deg \leq g} \oplus
\mathbb{C}[x]_{\deg = 2g+1}^{\text{monic}}$.
The maps $\Phi, \psi, \psi'$and $\phi$ are defined by
\begin{align}
  \label{Mum-NY-poly}  
  \Phi &: ~
  \begin{pmatrix}
    a(x) & b(x)\\
    c(x) & d(x) 
  \end{pmatrix}
  \mapsto
  \begin{pmatrix}
    \frac{a(x)-d(x)}{2} & c(x) \\
    b(x) & \frac{-a(x)+d(x)}{2}
  \end{pmatrix},
\\
  \psi &: l(x) \mapsto -\det l(x),
\\
  \psi' &: T(x) \mapsto ( \Tr T(x), -\det T(x) ),
\\
  \phi &: (f_1(x), f_2(x)) \mapsto
   f(x) = \bigl( \frac{1}{4} f_1(x)^2 + f_2(x) \bigr).
\end{align}
The commutativity of the diagram \eqref{diagI}
is readily seen.
\begin{proposition}
  \label{prop:T-flow}
  In the diagram \eqref{diagI}, we have the following:
\\
  (i) Via the map $\Phi,$
  the vector fields \eqref{Dg-flow}
  induce those defined in \eqref{time-evol}.
  In particular,   the vector fields
  $D_1, \cdots, D_g$ defined in \eqref{Dg-flow}
  are independent.
  \\
  (ii)
  Let $(f_1, f_2) \in \mathcal{C}^{(I)}$
  and let $f= \phi(f_1, f_2)$.
  The fiber $\psi^{' -1}(f_1, f_2)$ is isomorphic to the fiber 
  $\psi^{-1}(f) = \mathcal{M}_{g, f}.$
\end{proposition}
{\it Proof.}
(i) Let $L(x)= \Phi(T(x)).$
By using \eqref{Dg-flow}, we see that
the commuting vector fields generated by $D(x)$ induce
the flows on $\mathcal{M}_g$ as follows:
\begin{align}
  \label{flow-NY}
  \begin{split}
 & D(x_2) L(x_1)_{1,1}
  =
  \frac{1}{2} \Bigl(
                 \frac{L(x_1)_{1,2} L(x_2)_{2,1} 
                       - L(x_1)_{2,1} L(x_2)_{1,2}}{x_1-x_2} 
                 - L(x_1)_{2,1} L(x_2)_{2,1}
                 \Bigr)
  \\
  &D(x_2)L(x_1)_{2,1}
  = \frac{L(x_1)_{2,1} L(x_2)_{1,1} - L(x_1)_{1,1} L(x_2)_{2,1}}{x_1-x_2}
  \\
  &D(x_2) L(x_1)_{1,2}
  = \frac{L(x_1)_{1,1} L(x_2)_{1,2} - L(x_1)_{1,2} L(x_2)_{1,1}}{x_1-x_2} 
                     + L(x_1)_{1,1} L(x_2)_{2,1}
  \end{split}
\end{align}
By reference to  \eqref{Mum-NY-poly},
we see that the above flows induce the same vector field as that
generated by $D(x)$ \eqref{time-evol}.
\\
(ii) By the commutativity of \eqref{diagI},
the map $\Phi$ induces 
$\psi^{' -1}(f_1, f_2) \to \mathcal{M}_{g, f}.$
The inverse map of $\Phi$ is given by
\begin{align}
  \label{Phi-inverse}
l(x) = \begin{pmatrix} v(x) & w(x) \\ u(x) & -v(x) \end{pmatrix}
 \mapsto 
 \begin{pmatrix} 
 v(x)+\frac{1}{2}f_1(x) & u(x) \\ 
 w(x) & - v(x)+\frac{1}{2}f_1(x) 
 \end{pmatrix}. ~~~\square
\end{align}

Next we consider the system $\mathcal{D}_g^{(II)}$.
We define 
$\mathcal{C}^{(II)} \cong \mathbb{C}^{3g+3}$ 
to be the set of 
a pair $(f_1(x), f_2(x)) \in \mathbb{C}[x]^{\oplus 2}$
of polynomials of the form
\begin{align*}
  f_1(x) &= 2 x^{g+1} + f_1^{(1)} x^g + f_2^{(2)} x^{g-1} + \cdots 
           + f_1^{(g+1)}
  \\
 f_2(x) &= x^{2g+2} + f_2^{(1)} x^{2g+1} + f_2^{(2)} x^{2g} + \cdots
           + f_2^{(2g+2)}.
\end{align*}
We construct a commutative diagram similar to \eqref{diagI}
\begin{equation}
\label{diagII}
\begin{matrix}
\mathcal{D}_g^{(II)} & \supset \{ b_0 \not= 0 \} & 
   \overset{\Phi}{\longrightarrow} 
   & \mathcal{M}_g
\\[1mm]
{}_{\psi'} \downarrow & \quad   &  &  \downarrow_{\psi}
\\[1mm]
\mathcal{C}^{(II)} & \supset \{ f_1^{(1)} \not= f_2^{(1)} \} &
   \overset{\phi}{\longrightarrow}
   & \mathbb{C}[x]_{\deg = 2g+1}^{\text{monic}}. & 
\end{matrix}
\end{equation}
Here each maps are defined as follows:
\begin{align*}
  \Phi &: ~
  \begin{pmatrix}
    a(x) & b(x)\\
    c(x) & d(x) 
  \end{pmatrix}
  \mapsto
  \frac{1}{b_0}
  \begin{pmatrix}
    \frac{a(x)-d(x)}{2} & c(x) \\
    b(x) & \frac{-a(x)+d(x)}{2}
  \end{pmatrix},
\\
  \psi &: l(x) \mapsto -\det l(x),
\\
  \psi' &: T(x) \mapsto ( \Tr T(x), \det T(x) ),
\\
  \phi &: (f_1(x), f_2(x)) \mapsto
   f(x) = \frac{1}{f_1^{(1)}-f_2^{(1)}}
     \bigl( \frac{1}{4} f_1(x)^2 - f_2(x) \bigr).
\end{align*}
Note that 
$\Phi: \mathcal{D}_g^{(II)} \dasharrow \mathcal{M}_g$
and
$\phi: \mathcal{C}^{(II)} \dasharrow 
   \mathbb{C}[x]_{\deg = 2g+1}^{\text{monic}}$
are rational maps regular on open dense subsets indicated above.
The pull-back of $f_1^{(1)}-f_2^{(1)}$
by $\psi'$ is $b_0^2.$

\begin{proposition}
  \label{prop:T-flowII}
  In the diagram \eqref{diagII}, we have the following:
\\
  (i) Via the map $\Phi|_{\{ b_0 \not= 0 \}},$
  the vector fields \eqref{Dg-flow}
  induce those defined in \eqref{time-evol}.
  In particular, the vector fields
  $D_1, \cdots, D_g$ defined in \eqref{Dg-flow}
  are independent.
  \\
  (ii)
  Let $(f_1(x), f_2(x)) \in \mathcal{C}^{(II)}$
  be such that $f_1^{(1)} \not= f_2^{(1)}$,
  and let $f = \phi(f_1, f_2).$
  The fiber $\psi^{' -1}(f_1, f_2)$ is isomorphic to
  a disjoint union of two copies of
  $\psi^{-1}(f) = \mathcal{M}_{g, f}.$
\end{proposition}  
{\it Proof.}
(i) This is seen by the same way as Proposition \ref{prop:T-flow} (i).
\\
(ii) 
We take a complex number $b_0^*$ such that
$b_0^{* 2} = f_1^{(1)} - f_2^{(1)}.$
For $\sigma \in \{+1, -1 \},$
we see that $\Psi$ maps a subset
\[ \psi^{' -1}(f_1, f_2)^{\sigma}
  = \Bigl\{ \begin{pmatrix} a(x) & b(x) \\ c(x) & d(x) \end{pmatrix}
      \in \psi^{' -1}(f_1, f_2)
    ~|~ b_0 = \sigma b_0^* \Bigr\}
\]
of $\psi^{' -1}(f_1, f_2)$
isomorphically onto $\mathcal{M}_{g, f}$
with the inverse map
\[
   \mathcal{M}_{g, f} \to \psi^{' -1}(f_1, f_2)^{\sigma}
  ; ~
   \begin{pmatrix} v(x) & w(x) \\ u(x) & -v(x) \end{pmatrix}
   \mapsto
   \begin{pmatrix}
     \sigma b_0^* v + \frac{1}{2}f_1(x) & \sigma b_0^* u(x) \\
     \sigma b_0^* w(x) & -\sigma b_0^* v + \frac{1}{2}f_1(x)
   \end{pmatrix}.
\]
Since $\psi^{' -1}(f_1, f_2)$
is the disjoint union of
$\psi^{' -1}(f_1, f_2)^{+1}$ and $\psi^{' -1}(f_1, f_2)^{-1},$
the proposition follows.
~~ $\square$

Propositions \ref{prop:T-flow} and \ref{prop:T-flowII}
shows that for $i = I, II$ the generic level set of 
$\mathcal{D}_g^{(i)}$ is isomorphic to 
an affine part of an abelian variety of dimension $g$
on which the flows are linearized.
Thus we have shown the following.
\begin{theorem}
\label{int-D}
Both $\mathcal{D}_g^{(I)}$ and $\mathcal{D}_g^{(II)}$ are
algebraically completely integrable.
\end{theorem}


\subsection{Integrability of $\mathcal{Q}_g^{(I)}$ and 
$\mathcal{Q}_g^{(II)}$}

The works \cite{VeselovShabat93,Adler94,Takasaki03}
of Veselov-Shabat, Adler and Takasaki
give a concrete connection between 
the Noumi-Yamada systems \eqref{NY-I} \eqref{NY-II}
and the system $\mathcal{D}_g^{(i)}$ 
studied in the previous subsection.
We explain how these results imply the integrability 
of \eqref{NY-I} \eqref{NY-II} at $\alpha = 0$.

We begin with the Lax equation 
\cite{VeselovShabat93,Adler94}: 
\begin{align}
  \label{Laxform-NY}
  \frac{\partial l_k(x)}{\partial t} 
  = m_{k+1}(x+e_k-e_{k+1})\, l_k(x) - l_k(x) \,m_k(x)
\end{align}
where 
\begin{align*}
  &l_k(x) 
  = 
  \begin{pmatrix}
    \check{q}_k & 1 \\
    x + \check{q}_k^2 & \check{q}_k   
  \end{pmatrix}, 
  ~~
  m_k(x)
  = 
  \begin{pmatrix}
    0 & 1 \\
    x + \check{q}_k^2 - \dot{\check{q}}_k & 0
  \end{pmatrix},
\end{align*}
for $k=1,\ldots,N$. Here we fix the parameters $e_1, \ldots, e_{N}$,
and set the periodicity 
$\check q_{k+N} = \check q_k$ and $e_{k+N} = e_k$.
The Lax form \eqref{Laxform-NY} gives the evolution equation for $\check q_k$ 
called {\it the dressing chain}:
\begin{align}
  \label{dressing-chain}
  \frac{\partial {\check q}_k }{\partial t}
  + \frac{\partial {\check q}_{k+1}}{\partial t} 
  =
  \check q_{k+1}^2 - \check q_k^2 + e_k - e_{k+1}. 
\end{align}
According to the periodicity $N = 2g+1$ and $N = 2g+2$,
this is transformed into the $\alpha = 0$ limit of 
the Noumi-Yamada system of $A_{2g}^{(1)}$-type
\eqref{NY-I} and $A_{2g+1}^{(1)}$-type \eqref{NY-II}
by the variable transformation $q_k = \check q_k + \check q_{k+1}$.
This variable transformation gives an isomorphism 
between the phase spaces $\{(\check q_1,\check q_2,\ldots,\check q_N)\}$
and 
\begin{align*}
  &\mathcal{Q}_g^{(I)} = \{ q = (q_1, \ldots, q_{2g+1}) \}
  ~~\text{for $N=2g+1$},
  \\
  &\mathcal{Q}_g^{(II)} =
  \{ q = (q_1, \ldots, q_{2g+2}) 
  ~|~ \sum_{k=1}^{g+1} q_{2k} = \sum_{k=1}^{g+1} q_{2k-1}\}
  ~~\text{for $N=2g+2$}.
\end{align*}
Remark that an additional condition is required in the $N = 2g+2$ case.
We also use the notations $\mathcal{Q}_g^{(I)}$ and $\mathcal{Q}_g^{(II)}$
to denote the systems themselves.

We consider a diagram
\begin{align}
  \label{diag-QT}
  \begin{matrix}
\mathcal{Q}_g^{(i)} & \overset{\Lambda}{\longrightarrow} & \mathcal{D}_g^{(i)} \\
{}_{\psi''} \downarrow \quad & & ~ \downarrow_{\psi'} \\
\mathbb{C}^{g+1} & 
\overset{\lambda}{\longrightarrow} &
\mathcal{C}^{(i)} 
  \end{matrix}
\end{align}
for $i=I,II$.
The map $\Lambda$ is given by
\begin{align}
  \label{Lax-NY}
  (q_1, \ldots, q_{N}) \mapsto 
  T(x) = l_{N}(x-e_{N}) l_{N-1}(x-e_{N-1})\cdots l_1(x-e_1),
\end{align}
depending on $e_1, \ldots, e_{N}$.
The map $\psi''$ is given by 
\begin{align*}
  q &= (q_1, \cdots, q_N) 
  \mapsto
  h = (h_{\frac{1}{2}},h_{\frac{3}{2}},\ldots, h_{g+\frac{1}{2}}),
  \\
  &\text{where} 
  \Tr \Lambda(q) = 
  \begin{cases}
     h_{\frac{1}{2}} x^g + h_{\frac{3}{2}} x^{g-1} + \cdots + h_{g+\frac{1}{2}}
    \text{ for (I)},
    \\
    2 x^{g+1} + h_{\frac{1}{2}} x^g + h_{\frac{3}{2}} x^{g-1} 
    + \cdots + h_{g+\frac{1}{2}}
    \text{ for (II)}.
  \end{cases}
\end{align*}
The map $\lambda$ is defined with $e_k$'s so that the diagram is commutative:
\begin{align*}
  (h_{\frac{1}{2}},h_{\frac{3}{2}},\ldots, h_{g+\frac{1}{2}})
   \mapsto
   \begin{cases}
    (h_{\frac{1}{2}}x^g + h_{\frac{3}{2}}x^{g-1}+\cdots
           + h_{g+\frac{1}{2}}~,~ \prod_{k=1}^{2g+1} (x-e_k))
     \text{ for (I)},
     \\
    (2 x^{g+1} + h_{\frac{1}{2}}x^g + h_{\frac{3}{2}}x^{g-1}+\cdots
           + h_{g+\frac{1}{2}}~,~ -\prod_{k=1}^{2g+2} (x-e_k))     
     \text{ for (II)}.
   \end{cases}
\end{align*}

In \cite{VeselovShabat93,Takasaki03},
the $g$ commuting Hamiltonian vector fields 
on $\mathcal{Q}_g^{(i)}$ are constructed,
and their push-forward by $\Lambda$ are shown to coincide with
the vector fields $D_1, \ldots, D_g$
on $\mathcal{D}_g^{(i)}$ defined in \eqref{Dg-flow}.
(The push-forward of the vector field given by 
\eqref{NY-I} and \eqref{NY-II} is $D_1.$)
Moreover, Takasaki's  detailed study \cite{Takasaki03} 
on the map $\Lambda$ shows the following.
(The inverse of 
$\psi^{'' -1}(h) \to \psi^{' -1}(\lambda(h))$
is explicitly 
given by the map 
$\gamma_2$ in \cite{Takasaki03} eq.~(6.34). )

\begin{proposition}
  \label{prop:h}
  In the diagram \eqref{diag-QT}, 
  $\Lambda$ induces an isomorphism
  $\psi^{'' -1}(h) \overset{\sim}{\to} \psi^{' -1}(\lambda(h))$
  for any $h \in \mathbb{C}^{g+1}$.
\end{proposition}

By Proposition \ref{prop:h}, the integrability of 
the systems $\mathcal{D}_g^{(i)}$ shown in Theorem \ref{int-D}
comes to $\mathcal{Q}_g^{(i)}$ for $i=I$ and $II$.
\begin{theorem}
  \label{theorem:Qaci}
  Both $\mathcal{Q}_g^{(I)}$ and $\mathcal{Q}_g^{(II)}$ are
  algebraically completely integrable.
  In particular, 
  \\
  (i)
  the generic level set of $\mathcal{Q}_g^{(I)}$ is 
  isomorphic to the affine Jacobi variety
  $J(X) \setminus \Theta$.
  \\
  (ii)
  The generic level set of $\mathcal{Q}_g^{(II)}$ is 
  isomorphic to the disjoint union of two copies of $J(X) \setminus \Theta$.
\end{theorem}


\subsection{Solution of $\mathcal{Q}_g^{(I)}$ and $\mathcal{Q}_g^{(II)}$}
\label{subsec:NY-solution}

In the previous subsection, we have seen the integrability of 
$\mathcal{Q}_g^{(I)}$ and $\mathcal{Q}_g^{(II)}$.
Actually, we are in a far better position 
to deal with the solutions.
Our knowledge about the Mumford system is not just 
that it is integrable:
the solutions can be explicitly written 
in terms of {\it the theta functions},
at least for the coefficients of $u(x)$
\cite{Mumford-Book}. (See also Appendix A.)
This description, together with 
Proposition \ref{prop:T-flow}
enables us to write the time evolution of $q_i$
of the systems $\mathcal{Q}_g^{(I)}$ and $\mathcal{Q}_g^{(II)}$
as an algebraic function of the theta functions.
As a related work, we cite \cite{VeselovShabat93} where
the real solution for the 
dressing chain \eqref{dressing-chain} is discussed.

Let us see what happens in the cases of $g=1$.
The level set of the Mumford system $\mathcal{M}_{g,f}$ 
is the fiber of $\psi$ \eqref{Mum-phi} over 
$$
  f(x) = x^3 + f_1 x^2 + f_2 x + f_3.
$$
We assume $f(x)$ has no multiple zero,
and let $X$ be the elliptic curve defined by $y^2 = f(x)$.
Let $w_1$ and $w_2$ be the periods of the associated $\wp$-function.
Recall that on the tangent space of $\mathcal{M}_{1,f}$
we have one-dimensional vector field $\frac{\partial}{\partial t_1}$.
By working with the description of $u_1$ 
in terms of the theta functions \eqref{inverse-Abel},
we obtain $u_1$ and $v_{\frac{3}{2}}$ 
in terms of the Weiestra\ss ~$\wp$-function:
\begin{align}
  \label{M1-solution}
  u_1 = -\wp(\text{\scriptsize{$\frac{t_1 + \tau}{2}$}}),
  ~~ 
  v_{\frac{3}{2}} = -\text{\scriptsize{$\frac{\partial}{\partial t_1}$}} 
                    \wp(\text{\scriptsize{$\frac{t_1 + \tau}{2}$}}).
\end{align}
Here $\tau$ is a constant which is uniquely determined by
the initial data $q |_{t_1=0} \in \mathcal{Q}_1^{(i)}$ for $i=I,II$.
In the following we write $\wp = \wp(\frac{t_1 + \tau}{2})$ 
and $\partial_{t_1} \wp = \frac{\partial}{\partial t_1} \wp$.
\\[2mm]
{\it Solution of $\mathcal{Q}_1^{(I)}$:}
We have three dynamical variables $q_1, q_2$ and $q_3$.
We fix the integrals of motion $h_{\frac{1}{2}}, h_{\frac{3}{2}}$
and the parameters $e_1, e_2, e_3$.
Then the curve $X$ is determined by 
$$
  f(x) = \frac{1}{4} (h_{\frac{1}{2}} x + h_{\frac{3}{2}})^2
         + (x-e_1)(x-e_2)(x-e_3).
$$
The map $\psi^{''}$ and the isomorphism 
$\psi^{'' -1}(h) \simeq \mathcal{M}_{1,f}$
induced by $\Phi \circ \Lambda$ 
give the following relations 
(see the diagrams \eqref{diagI} and \eqref{diag-QT}):
\begin{align}
  \label{QIrelations}
  \begin{split}
  &h_{\frac{1}{2}} = q_1 + q_2 + q_3, 
  \\
  &u_1 = -e_2 + q_1 q_2, 
  ~~
  v_{\frac{3}{2}} = - q_2(u_1+e_1) 
                     +\frac{1}{2}(h_{\frac{1}{2}} u_1 - h_{\frac{3}{2}}). 
  \end{split}
\end{align}
The relations \eqref{M1-solution} and \eqref{QIrelations}
explicitly give the solution of $q_i$'s
as functions of $t_1$:
\begin{align*}
  &q_1(t_1) = \frac{-\partial_{t_1} \wp 
                    + h_{\frac{1}{2}} \wp + h_{\frac{3}{2}}}
                   {2(\wp - e_1)},
  ~~
  q_2(t_1) = - \frac{2(\wp - e_1) (\wp - e_2)}
               {-\partial_{t_1} \wp + h_{\frac{1}{2}} \wp + h_{\frac{3}{2}}},
  \\
  &q_3(t_1) = -\frac{h_{\frac{1}{2}} e_1 + h_{\frac{3}{2}}}{\wp - e_1}
        + \frac{2(\wp - e_2) (e_3 - e_1)}
               {-\partial_{t_1} \wp + h_{\frac{1}{2}} \wp + h_{\frac{3}{2}}}.
\end{align*}
Here $\tau$ is uniquely determined by the initial data
$q|_{t_1=0} = (q_1(0),q_2(0),q_3(0))$.
\\[2mm] 
{\it Solution of $\mathcal{Q}_1^{(II)}$:}
In this case we have four dynamical variables $q_1,\ldots, q_4$.
We fix the integrals of motion $b_0,h_{\frac{1}{2}}$ and $h_{\frac{3}{2}}$,
and four parameters $e_1,\ldots,e_4$.
Then the curve $X$ is determined by
$$
  f(x) = \frac{1}{b_0^2} \bigl( 
         \frac{1}{4} (2 x^2 + h_{\frac{1}{2}} x + h_{\frac{3}{2}})^2
         - (x-e_1)(x-e_2)(x-e_3)(x-e_4)
         \bigr),
$$
which turns out to be a monic cubic polynomial of $x$.
The composition $\Phi \circ \Lambda$ of maps
in the diagrams \eqref{diagII} and \eqref{diag-QT} 
induces the following relations
\begin{align}
  \label{QIIrelations}
  \begin{split}
  &b_0 = q_1 + q_3 = q_2 + q_4,
  \\
  &u_1 = \frac{1}{b_0}(- e_2 q_3 - e_3 q_1 + q_1 q_2 q_3),
  ~~
  v_{\frac{3}{2}} = \frac{1}{b_0} \bigl(
                     u_1 (e_3 - q_2 q_3) 
                     + \frac{1}{2}(h_{\frac{1}{2}} u_1 - h_{\frac{3}{2}})
                     \bigr).
  \end{split}           
\end{align}
The relations \eqref{M1-solution} and \eqref{QIIrelations}
explicitly give the solution for $q_i$'s as 
\begin{align*}
  &q_1(t_1) =\frac{2 \wp b_0 (e_2-\wp)}
             {2 e_2 \wp -2b_0 \wp' + \wp h_{\frac{1}{2}} + h_{\frac{3}{2}}},
  \\
  &q_2(t_1) 
     = 
     \frac{(2 e_2 \wp -2b_0 \wp' + \wp h_{\frac{1}{2}} + h_{\frac{3}{2}})
               (2 e_3 \wp -2b_0 \wp' + \wp h_{\frac{1}{2}} + h_{\frac{3}{2}})
               }
              {2 \wp b_0 
               (-2b_0 \wp'+ 2 \wp^2 + \wp h_{\frac{1}{2}} + h_{\frac{3}{2}})},
  \\
  &q_3(t_1) = b_0 - q_1(t_1), ~~~ q_4(t_1) = b_0 - q_2(t_1).
\end{align*}
Here $\tau$ is determined by $q|_{t_1 =0}$ as the (I) case.
 
In general $g$ case, we fix $e_i$'s and a fiber of $\psi''$ over 
$\mathbb{C}^{g+1}$, 
and let $X$ be the hyperelliptic curve given by 
\begin{align*}
 y^2 = f(x) = 
 \begin{cases}
   \displaystyle{
        \frac{1}{4}(h_{\frac{1}{2}} x^g + \cdots + h_{g+\frac{1}{2}})^2 
        + \prod_{i=1}^{2g+1}(x-e_i)
   }
    ~~ \text{for (I)},
 \\
  \displaystyle{
       \frac{1}{b_0^2} \bigl( 
       \frac{1}{4}(2 x^{g+1} + h_{\frac{1}{2}} x^g 
                   + \cdots + h_{g+\frac{1}{2}})^2 
         - \prod_{i=1}^{2g+2}(x-e_i)
       \bigr)
   }
   ~~ \text{for (II)}.
 \end{cases} 
\end{align*}
By making use of the birational map $\Phi \circ \Lambda$
and the description of $u_i$ 
in terms of the theta functions \eqref{inverse-Abel},
we can obtain an explicit description of the solution for 
$\mathcal{Q}_g^{(I)}$ and $\mathcal{Q}_g^{(II)}$.


\setcounter{equation}{0}
\setcounter{theorem}{0}

\renewcommand{\theequation}{A.\arabic{equation}}
\renewcommand{\thetheorem}{A.\arabic{theorem}}

\subsection*{Appendix A ~ Theta functions}

Following \cite{Mumford-Book} we first recall 
the definition and some properties for the theta functions.
Let $X$ be a smooth hyperelliptic curve of genus $g$
given by $y^2 = f(x)$ in \S 3.1.
Let $a_1, a_2, \ldots, a_{2g+1}$ and  $\infty$
be the branch points,
and let $B=\{1,2,\ldots,2g+1\}$.
The theta function $\theta(\vec z, \Omega)$ of integer characteristics 
is defined as
\begin{align*}
  \theta(\vec z, \Omega) = \sum_{\vec n \in \mathbb{Z}^g} 
                    \exp(\pi i ^t \vec n \Omega \vec n 
                         + 2 \pi i ^t \vec n \cdot \vec z),
\end{align*}
where $\vec n$ and $\vec z$ are column vectors whose length are $g$,
and $\Omega$ is the $g$ by $g$ period matrix of $X$.
Especially $\vec z$ gives the coordinate on 
$J(X) \simeq \mathbb{C}^g / (\mathbb{Z} + \Omega \mathbb{Z})$.
The theta functions with rational characteristics is 
\begin{align*}
  \theta\Bigl[
    \begin{matrix}
    \vec a \\ \vec b
    \end{matrix}
    \Bigr](\vec z)
  =
  \exp\bigl(\pi i ^t \vec a \Omega \vec a 
      + 2 \pi i ^t \vec a \cdot (\vec z + \vec b)\bigr)
  \cdot \theta(\vec z + \Omega \vec a + \vec b, \Omega), ~~
  \vec a, ~ \vec b \in \mathbb{Q}^g.
\end{align*}
Each branch point $a_k$ is related to 
one of the half periods of $J(X)$ by the
integration of the holomorphic one-forms 
on $X$ as 
$\int_{\infty}^{a_k} \vec w = \Omega \vec\eta_k + \vec\eta_k^\prime$
for $k \in B$.
Let $\Delta$ be a summation of the half periods 
$\sum_{k=1}^{g+1} 
 \bigl( \Omega \vec \eta_{2k-1} + \vec \eta_{2k-1}^\prime \bigr)$.

We introduce an important identity for the theta functions:
\begin{proposition}
  \label{theta-id}
  (\cite{Mumford-Book} Corollary 7.5)
  There exists a subset $V$ of $B$ of cardinal $g+1$
  such that the functions on $J(X)$
  \begin{align*}
    \lambda_{k}(\vec z)
    = 
    \Bigl(
     \frac{\theta\text{\Kakko{\eta_k}{\eta_k^\prime}}(0)~
            \theta\text{\Kakko{\eta_k}{\eta_k^\prime}}(\vec z + \vec \Delta)}
           {\theta(0) ~\theta(\vec z + \vec \Delta)}
    \Bigr)^2,~~ \text{for } k \in V, 
  \end{align*}
  satisfy $\displaystyle{\sum_{k \in V}} \lambda_k(\vec z) = 1$.
\end{proposition}

Let $\xi$ be the isomorphism of Theorem \ref{oddMum-Jac} 
$\xi : \mathcal{M}_{g,f} \overset{\sim}{\rightarrow} J(X)\setminus \Theta;
~l(x) \mapsto \vec z$.
The following theorem describes $u(x)$ explicitly in terms of $\vec z$:
\begin{theorem}(\cite{Mumford-Book} \S 5 Theorem 5.3, Proposition 5.10)
  (i)
  Let $V$ be the subset of $B$ defined in Proposition \ref{theta-id}.
  For each element $l(x)$ of $\mathcal{M}_{g,f}$ such that 
  $\xi(l(x)) = \vec z$,
  the coefficients of $u(x)=u(\vec z)(x)$ are written as
  \begin{align}
    \label{inverse-Abel}
    u(x) = \sum_{k \in V} \bigl( \lambda_{k}(\vec z) 
             \prod_{j \in V,j\neq k} (x-x(a_j)) \bigr),
\end{align}
Here $x(a_j)$ denotes the $x$-coordinate of the branch point $a_j$.
\\
  (ii)
  The coordinate on $J(X)$ given by $\vec z = (z_1,\ldots,z_g)$ 
  linearizes the vector field $D(x)$ \eqref{time-evol}, namely
  we can write each $D_i = \frac{\partial}{\partial t_i}$ as 
  a linear combination of $\frac{\partial}{\partial z_i}$.
\end{theorem}
Note that the r.~h.~s. of \eqref{inverse-Abel} becomes monic in $x$ 
due to Proposition \ref{theta-id}.


\end{document}